\documentclass[reprint,superscriptaddress,amsmath,amssymb,aps,prb,longbibliography]{revtex4-2}
\usepackage{graphics,graphicx,epsfig,color,latexsym,float,amsmath,bm}
\usepackage[colorlinks,citecolor=blue,linkcolor=blue,urlcolor=blue]{hyperref}
\usepackage[version=4]{mhchem}
\usepackage{wrapfig}
\usepackage{ulem}
\usepackage{lipsum}
\usepackage{placeins}
\usepackage{braket}
\usepackage{appendix}

\def\thefootnote{$^\dagger$}\footnotetext{These authors contributed equally to this work.}

\begin{document}

\title{Engineering many-body quantum Hamiltonians with non-ergodic properties using quantum Monte Carlo}

\author{Nyayabanta Swain\thefootnote{}}
\affiliation{Centre for Advanced 2D Materials, National University of Singapore, 6 Science Drive 2, Singapore 117546}
\affiliation{Department of Materials Science and Engineering, National University of Singapore, 9 Engineering Drive 1, Singapore 117575}

\author{Ho-Kin Tang\thefootnote{}}
\affiliation{Centre for Advanced 2D Materials, National University of Singapore, 6 Science Drive 2, Singapore 117546}
\affiliation{School of Science, Harbin Institute of Technology, Shenzhen, P. R. China 518055}

\author{Darryl Chuan Wei Foo}
\affiliation{Centre for Advanced 2D Materials, National University of Singapore, 6 Science Drive 2, Singapore 117546}

\author{Brian J. J. Khor}
\affiliation{Centre for Advanced 2D Materials, National University of Singapore, 6 Science Drive 2, Singapore 117546}

\author{\mbox{Gabriel Lemari\'e}}
\affiliation{MajuLab, CNRS-UCA-SU-NUS-NTU International Joint Research Unit IRL 3654, Singapore}
\affiliation{Centre for Quantum Technologies, National University of Singapore, Singapore 117543}
\affiliation{Laboratoire de Physique Th\'{e}orique, Universit\'{e} de Toulouse, CNRS, UPS, France}

\author{Fakher F. Assaad}
\affiliation{Institut f\"ur Theoretische Physik und Astrophysik, Universit\"at W\"urzburg-Dresden Cluster of Excellence ct.qmat, Universit\"at W\"urzburg, Am Hubland, D-97074 W\"urzburg, Germany.}

\author{Pinaki Sengupta}
\affiliation{Centre for Advanced 2D Materials, National University of Singapore, 6 Science Drive 2, Singapore 117546}
\affiliation{School of Physical and Mathematical Sciences, Nanyang Technological University, 21 Nanyang Link, Singapore 637371}

\author{Shaffique Adam}
\email{Corresponding author. Email: shaffique@wustl.edu}
\affiliation{Centre for Advanced 2D Materials, National University of Singapore, 6 Science Drive 2, Singapore 117546}
\affiliation{Department of Materials Science and Engineering, National University of Singapore, 9 Engineering Drive 1, Singapore 117575}
\affiliation{Yale-NUS College, 16 College Ave West, Singapore 138527}
\affiliation{Department of Physics, Washington University in St. Louis, St. Louis, Missouri 63130, United States}

\date{\today}

\begin{abstract}
We present a computational framework to identify Hamiltonians of interacting quantum many-body systems that host non-ergodic excited states. We combine quantum Monte Carlo simulations with the recently proposed eigenstate-to-Hamiltonian construction, which maps the ground state of a specified parent Hamiltonian to a single non-ergodic excited state of a new derived Hamiltonian. This engineered Hamiltonian contains non-trivial, systematically-obtained, and emergent features that are responsible for its non-ergodic properties.  We demonstrate this approach by applying it to quantum many-body scar states where we discover a previously unreported family of Hamiltonians with spatially oscillating spin exchange couplings that host scar-like properties, including revivals in the quantum dynamics, and towers in the inverse participation ratio; and to many-body localization, where we find a two-dimensional Hamiltonian with correlated disorder that exhibits non-ergodic scaling of the participation entropy and inverse participation ratios of order unity.  The method can be applied to other known ground states to discover new quantum many-body systems with non-ergodic excited states.
\end{abstract}

\maketitle

\section{Introduction}

Statistical mechanics is foundational to classical many-body theory, underpinning our understanding of the natural world and responsible for virtually all industrial progress before the advent of the transistor.  Concepts from statistical mechanics have attained such familiarity that we describe other areas of physics using the same language. For example, path integral formulations of quantum mechanics invoke the unphysical concept of imaginary time to recast the propagator as a Boltzmann factor, and the normalization constant as a partition function. This success at transplanting the ideas, formalism and language of statistical mechanics to quantum theory has led to the assumption that the postulates of statistical mechanics continue to hold.  This is the eigenstate thermalization hypothesis \cite{ETH_1991,ETH_1994,ETH_review_2016} that can be understood as follows.  For an ergodic many-body quantum system, a state evolves under unitary time evolution, remaining pure and thus, non-thermal. However, after a partial trace is done over a designated subsystem, the remaining subsystem density matrix is mixed, thermal, and for sufficiently long time and large system size independent of initial conditions. In other words, an ergodic quantum many-body system will act as its own bath and thermalize in the usual statistical physics sense. 

Recently, several counter-examples to this ergodic scenario have been found in both theory and experiment that show non-ergodic properties including for integrable systems~\cite{ETH_review_2016}, quantum many-body scars (QMBS)~\cite{rydberg_atom_2017,scar_states_Turner2018,WenWei_2019,PhysRevLett.122.173401,scar_states_Serbyn2021,rev_scars_2023}, many-body localization (MBL)~\cite{Nandkishore2014,Alet2018,Abanin2019,GOPALAKRISHNAN20201,sierant2024manybody},
and systems with fragmented Hilbert spaces~\cite{Frey2022,Nicolau2023,Andreadakis2023,Kochergin2023}.  Non-ergodicity can be characterized by the scaling of the entanglement entropy $S_\mathrm{E}=-\mathrm{Tr}\rho_\mathrm{A}\log\rho_\mathrm{A}$, with $\rho_\mathrm{A}$ the density matrix for subsystem A.  Thermal states are ergodic and have $S_\mathrm{E}$ proportional to the volume of A, while non-ergodic states have $S_\mathrm{E}$ growing more slowly, for example with the area of A \cite{Bauer_2013} or the log of the volume of A \cite{PhysRevB.98.235156}. Non-ergodic systems can preserve quantum information and present the same richness to quantum systems that non-fully chaotic classical dynamics like the solar system give to classical systems.  Without tools from statistical mechanics, we lack a theoretical framework to understand the emergence of these non-ergodic properties.

From another perspective, non-ergodic states are as familiar as thermal states. Most ground states of quantum many-body systems are generically non-ergodic and follow an area-law scaling of their entanglement entropy \cite{RevModPhys.82.277}. The description of ground state properties is much better developed than for excited states, and in particular, numerical approaches such as quantum Monte Carlo (QMC) can handle large Hilbert space sizes $\gtrsim 2^{1000}$, and can be applied to different types of many-body ground states.  In this work we demonstrate how to apply the QMC approach developed for low-energy properties of quantum many-body systems to understand non-ergodic properties at high energy.  We find new Hamiltonians with emergent symmetries that support non-ergodic states including 
Hamiltonians with oscillating spin exchange couplings that have QMBS states and two-dimensional correlated disorder Hamiltonians with many-body localization.  These symmetry properties emerge non-trivially and systematically from the method and were not known previously.

\begin{figure*}[t]
\centerline{
\includegraphics[width=0.9\linewidth]{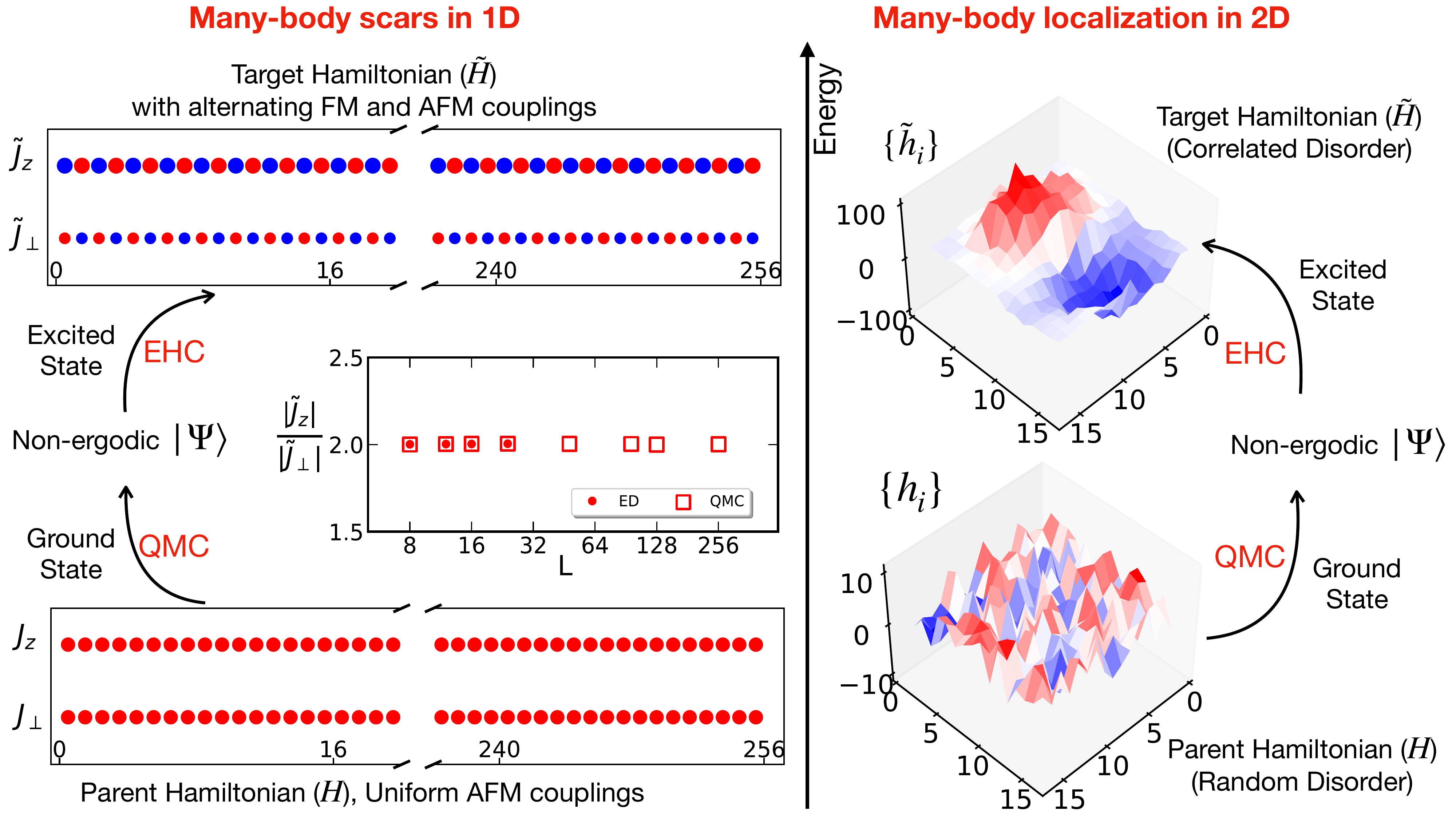}
}
\caption{\textbf{Generating Hamiltonians with non-ergodic properties using the Eigenstate to Hamiltonian Construction combined with the Quantum Monte Carlo method:}
This figure illustrates our EHC-QMC framework for studying quantum many-body scars in 1D (left) and many-body localization in 2D (right). Starting from a parent Hamiltonian $H$: a homogeneous, antiferromagnetic Heisenberg spin-1/2 chain for QMBS (bottom left) and a 2D Heisenberg model in a random magnetic field for MBL (bottom right), QMC is first used to compute the ground state $\ket{\Psi}$ of $H$ that is inherently non-ergodic.  In each case, the EHC procedure identifies a new target Hamiltonian $\tilde{H}$ (upper panels) that host 
$\ket{\Psi}$ as an approximate highly excited eigenstate.  We find that for QMBS the target Hamiltonian has alternating ferromagnetic (red) and antiferromagnetic (blue) spin-flip (top layer) and spin-aligned (bottom layer) exchange couplings, an emergent symmetry responsible for the scar properties.  For MBL the new target Hamiltonian (top right) has a strong emergent spatial correlation of the disorder.} 
\label{fig:1}
\end{figure*} 

One may wonder how QMC, whose sampling is ergodic in nature, traversing the full Hilbert space could possibly give results for a non-ergodic system.  While it is true that QMC sampling generates any possible state in the Hilbert space, states are then accepted or rejected according to some criteria, such as energy in the Metropolis-Hastings algorithm.  Rejected states no longer contribute to the expectation value computation.  Since a finite sized system has a finite gap between the ground state and first excited state, an appropriate choice of ``temperature'' in the Boltzmann weighting function then ensures the exponentially vanishing contribution of any excited state to the expectation value.  This is what allows QMC to accurately quantify the observables for non-ergodic ground states.  While QMC uses the Boltzmann factor as a weighting function, this does not imply that the method implicitly assumes the eigenstate thermalization hypothesis.  Here the Boltzmann factor is used merely as a weighting function to target the ground state properties at sufficiently small temperature. In principle, one could use a different weighting function that is unrelated to thermalization and statistical mechanics.  One could use a small negative temperature to target only the highest energy eigenstate, or in particular, one could use the shift-invert method \cite{Alet2018_shift-invert} with the weighting function $(E - H)^{-1}$, where $H$ is the Hamiltonian, to specifically target the highly excited state with energy closest to $E$.  While the use of the shift-invert operator might be the first choice, at least conceptually, for targeting excited states using QMC, in practice this does not work for our purpose.  This is because the form of the shift-invert function introduces competing frustrated interactions resulting in a sign problem for the QMC.  Instead, in this work, we show that combining the eigenstate-to-Hamiltonian construction (EHC)~\cite{Chertkov2018,Qi2019} with QMC successfully obtains Hamiltonians with non-ergodic excited state properties.


\section{EHC-QMC: Eigenstate to Hamiltonian Construction with Quantum Monte Carlo}

While studies of quantum systems typically commence with Hamiltonians from which eigenstates or various observables are derived, the EHC~\cite{Chertkov2018,Qi2019,Dupont2019a} offers an alternative by addressing the reverse question: given a particular eigenstate, what Hamiltonian hosts it?  Figure 1 illustrates our application of the EHC. It involves first defining an input state $\ket{\Psi}$, assumed here as a ground state of a parent Hamiltonian $H$, and a set of local operators $\{\mathcal O_i\}_{i=1,N_\mathcal O}$, such that $H=\sum_{i=1}^{N_\mathcal O} h_i \mathcal O_i$. The next step is to calculate the covariance matrix $\mathcal C$ whose elements are
\begin{equation}
\mathcal C_{ij}=\bra{\Psi} \mathcal O_i \mathcal O_j\ket{\Psi}-\bra{\Psi} \mathcal O_i\ket{\Psi}\bra{\Psi} \mathcal O_j\ket{\Psi}
\end{equation}
An eigenvector of the covariance matrix $\mathcal C$ provides the coefficients $\tilde{h}_i$ defining a target Hamiltonian $\tilde{H}=\sum_{i=1}^{N_\mathcal O} \tilde{h}_i \mathcal O_i$. Its associated eigenvalue represents the energy variance of $\ket{\Psi}$ with respect to the target Hamiltonian $\tilde{H}$. If this eigenvalue is zero, then $\ket{\Psi}$ is an exact eigenstate of $\tilde{H}$. Since $\ket{\Psi}$ is the ground state of a local Hamiltonian, it satisfies an area law. But $\ket{\Psi}$ is also an excited eigenstate of $\tilde{H}$ and therefore violates the ETH by construction. 

The EHC was originally conceived  to identify new Hamiltonians sharing the same ground state as the parent Hamiltonian and considered only zero eigenvalues of the covariance matrix where the mapping between the ground states of the parent and target Hamiltonians is exact \cite{Chertkov2018}.  By contrast, in this work we use this approach as an approximate method to find a target Hamiltonian $\tilde H$ with $\ket{\Psi}$ as a highly excited state.  Here, the eigenvalue of the covariance matrix considered is non-zero, but vanishingly small in the thermodynamic limit.  Our approach is similar to Ref.~\cite{Dupont2019a} where the authors have used the density matrix renormalization group (DMRG) numerical method in 1D supplemented by the EHC to approximately map many-body localization to a class of localized ground states known as the Bose-glass.  

The effectiveness of the EHC relies on the judicious choice of the set of operators $\{\mathcal O_i\}_{i=1,N_\mathcal O}$. The calculation of the elements $\mathcal C_{ij}$ can be more involved depending on this choice, but the size of the covariance matrix $\mathcal C$ depends on the number of these operators $N_{\mathcal O}$ (rather than on the dimension of the Hilbert space which is exponential in system size). With a proper choice, the method can be highly efficient, enabling exploration of large system sizes and overcoming the limitations of exact approaches.  Further details of our EHC procedure is described in the Appendix \ref{appendix_A}.   

Developing approximate but reliable methods to describe the high energy properties of quantum many body systems, that are challenging to access by exact approaches, is a crucial goal in this field \cite{Khemani2016,Yu2017,nonergodicdmrg1,nonergodicdmrg4}. Our approach enables us to harness QMC's ability to access larger system sizes to explore non-ergodic excited states.  We note that nearly all previously known models with non-ergodic properties were constructed through inspiration, fine-tuning, or serendipity \cite{Oganesyan2007,Pal2010,longrangembl1,Stark-MBL,starkmbl2,Stark-MBL_Doggen,starkmbl3}. Our method circumvents this by providing a systematic approach to extend non-ergodic ground state properties to the high-energy sector. 
In addition, we have developed metrics to quantify the accuracy of the approximation providing diagnostics to determine when the EHC-QMC construction works.

Before we describe our results, we note some drawbacks of our approach.  First, QMC is unable to obtain the full spectrum of the target Hamiltonian.  It is therefore necessary to complement our approach with exact diagonalization (ED) at smaller system sizes to fully characterize the properties of the non-ergodic states we have identified.  For example, we use ED to confirm that the ground state maps to a single eigenstate as opposed to a superposition of excited states.  Second, the QMC-EHC approach is non-exhaustive by construction.  We are unable to access non-ergodic excited states with a large average sign $\sim 1$ as these will suffer from a sign problem~\cite{Theveniaut2019}.  Similarly, we are unable to access ergodic excited states, and therefore unable to characterize the full ergodic-to-non-ergodic transition of excited states as would be necessary to describe a many-body localization transition. Despite these drawbacks, our approach extends the scope of previous EHC studies~\cite{Dupont2019a} by accessing substantially larger system sizes, and also higher dimensions (e.g. 2D) to probe essential features of Hamiltonians that support non-ergodic properties.


While we illustrate the method by focusing on two main classes of non-ergodic excited states, namely, QMBS and MBL, there are some key differences between them:  For MBL, all the excited states (at least within some energy window) are non-ergodic, while the non-ergodic QBMS states differ from the other excited eigenstates that are ergodic.  Moreover, our MBL non-ergodic states occurs in a disordered system that necessitates appropriately averaging over hundreds of disorder configurations, while the QMBS states are for a single realization of a clean system.  Despite these differences we are still able to use the same EHC-QMC framework to discover new examples in both these cases.

\section{Quantum Many Body Scar States}

\begin{figure*}[t]
\centerline{
\includegraphics[width=0.95\linewidth]{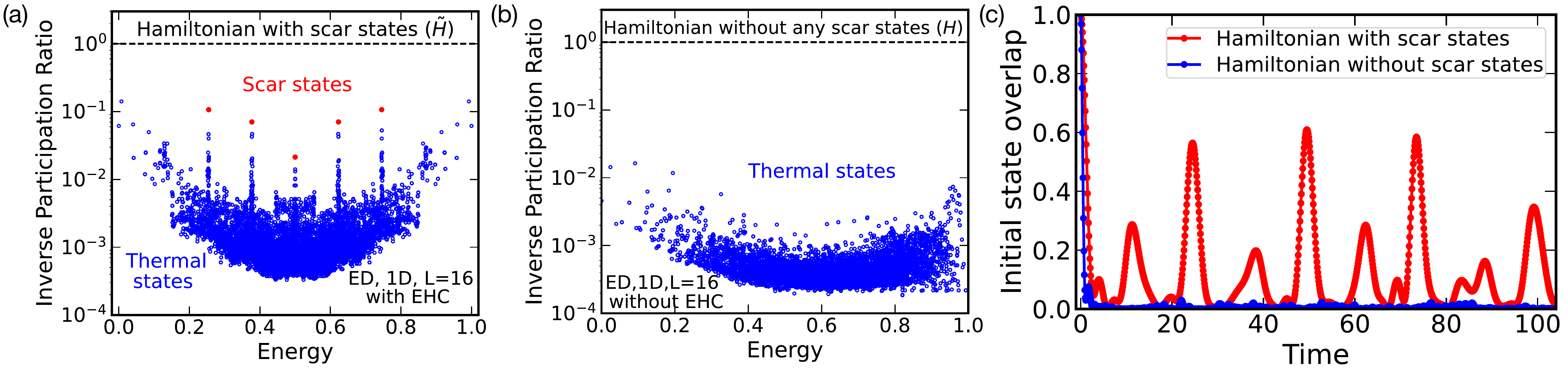}
}
\caption{
{\bf Characteristics of the scar model:}
(a)-(b)~Study of the non-ergodic properties of individual eigenstates of two different Hamiltonians ($H$ and $\tilde{H}$) of 1D chains of length $L=16$ with periodic boundary conditions. The Hamiltonian with scar states (parameterized by nearest neighbour couplings $|\tilde{J}_{\perp}| = 0.45J$ and $|\tilde{J}_{z}| = 0.9J$, see Eq.~\ref{eq1_mbs}), hosts towers of states with larger inverse participation ratios in the excitation energy spectrum than the typical thermal states. For illustration, we have highlighted the states at the top of the towers in the IPR spectrum in red. These states exhibit anomalously large non-ergodic properties as compared to the typical thermal states. The Hamiltonian $H$ without any scar states is parameterized by couplings $J_{\perp} = J_{z} = J$ with no change in the sign of the coupling has IPR values typical of thermal states. The excitation energy values are rescaled such that $E_{min} =0$ and $E_{max} =1$. (c)~Variation of the overlap amplitude of the initial state with its time evolved version as a function of time (in units of $1/J$).  The Hamiltonian in (a) is shown in red and displays a revival of the initial wavefunction that is a characteristic of scar-like behavior, while the Hamiltonian in (b) shown in blue has no such revivals.} 
\label{fig:2}
\end{figure*}

The usual behaviour for most Hamiltonians is that all highly excited states are thermal/ergodic, with, for example, an entanglement entropy scaling with volume. Yet Hamiltonians have been discovered both theoretically and experimentally~\cite{scar_states_Turner2018,scar_states_Serbyn2021,rev_scars_2023,rydberg_atom_2017}
in which a vanishing fraction of states are the so-called ``scar states'', with anomalous non-ergodic properties, e.g. sub-volume entropy scaling. If an initial state has a strong overlap with those scar states, the time-evolved state will exhibit periodic revivals \cite{scar_states_Turner2018}, in strong contrast with the systematic relaxation to an effective thermal equilibrium for systems with ergodic properties \cite{ETH_1991,ETH_1994,ETH_review_2016}.  These scar states are named for an analogy to classical scars, unstable periodic orbits in systems that do not generally host periodic orbits, and to our knowledge, the engineering of systems hosting such scar states to date has required fine-tuned kinetic constraints that emulate the formation of these classical closed orbits~\cite{scar_states_Serbyn2021,PhysRevLett.123.030601}. The QMC-EHC approach therefore provides an untapped niche for the systematic development of scar Hamiltonians.  We show here that starting from the ground state of a clean, transitionally invariant, spin chain model, the EHC provides a target Hamiltonian with quantum many-body scarring. 
We confine our search of target Hamiltonians to the possibly inhomogenous spin-$1/2$ Heisenberg model, on a 1D chain with periodic boundary conditions
\begin{equation}
\tilde{H} = \sum_{i} [ \tilde{J}_{\perp,i} \tfrac{1}{2}(S_i^{+}  S_{i+1}^{-} + S_i^{-}  S_{i+1}^{+}) + \tilde{J}_{z,i} S_i^z S_{i+1}^z ] 
\label{eq1_mbs}
\end{equation}
where $S^{+(-)}_i$ is the raising (lowering) operator for a spin at site $i$, $S^z_i$ is the projection of that spin on the $z$-axis, $\tilde{J}_{\perp,i}$ are the couplings for swaps of nearest neighbour singlets and $\tilde{J}_{z,i}$ is the site-resolved nearest neighbour Ising interaction. The family of Hamiltonians of Eqn.~\ref{eq1_mbs} commutes with the total spin operator, with the largest sector of net zero spin having a Hilbert space size of $\frac{N!}{((N/2)!)^2}\sim N^{-1/2}2^N$, exponential in the system size $N$. We use the ground state of the \textit{homogeneous} case, $\tilde{J}_{\perp,i}=\tilde{J}_{z,i}=J=1$, as our input state $|\Psi\rangle$. This state displays antiferromagnetic quasi-long-range order and the usual non-ergodic properties associated with ground states \cite{Affleck_1998}. We choose the EHC operator basis as ${\mathcal O}_i\equiv \tfrac{1}{2}(S_i^{+} S_{i+1}^{-} + S_i^{-} S_{i+1}^{+})$ and ${\mathcal O}_{i+N}=S_i^z S_{i+1}^z$ with $i=1,\ldots,N$, i.e., the number of operators $N_{\mathcal O}=2N$, to construct the $2N \times 2N$ covariance matrix, ${\mathcal C}$, and perform the EHC as described above. 

We note some important aspects of our construction of target Hamiltonians with scar properties. First and unexpectedly, not all target Hamiltonians defined by the eigenvectors of the covariance matrix exhibit QMBS.  Rather, we find scar states for the smallest nontrivial and nondegenerate eigenvalue of the covariance matrix. This eigenvalue vanishes in the thermodynamic limit, (see section \ref{sec5}). We have systematically checked that this choice consistently yields target Hamiltonians with scarring properties.  Second, while the covariance matrix should strictly obey the translation invariance of the parent Hamiltonian, i.e., correlators $\langle {\mathcal O}_i {\mathcal O}_j \rangle$ depend only on $i-j$ (and this property is straightforwardly checked using ED), QMC provides statistical estimations of these correlators which imperfectly fulfill this constraint.  We therefore impose the constraint by replacing $\langle {\mathcal O}_i {\mathcal O}_{i+r} \rangle \equiv \sum_i \langle {\mathcal O}_i {\mathcal O}_{i+r} \rangle / N $.  In this way for the small eigenvalue regime of the covariance matrix that we are focusing on, both ED and QMC results agree perfectly for small system sizes. The extension to large system sizes enabled by QMC fits well with the results at small sizes, as shown in the middle inset of Fig.~\ref{fig:1}.

The target Hamiltonian obtained using this method is illustrated in Fig.~\ref{fig:1}, where the couplings have equal magnitude but alternating signs at every bond, with the sign of $\tilde{J}_{\perp,i}$ opposite to that of $\tilde{J}_{z,i}$ and the magnitudes locked in the ratio $|\tilde{J}_{\perp,i}/\tilde{J}_{z,i}|=1/2$ at each bond. To our knowledge this is the first time such a Hamiltonian has been identified as a host for scar states.  Figures \ref{fig:2}(a) and (b) demonstrate the QMBS properties. In (a), we observe the familiar scar towers of non-ergodic properties reminiscent of the PXP model \cite{PXP_2008}.  We constrast this with panel (b) where we show the case of the homogeneous Heisenberg model, which does not host any scar states. In these figures, we represent the value of the Inverse Participation Ratio (IPR) as a measure of the inverse volume occupied by an eigenstate $\ket{\psi_\alpha}$ of the target Hamiltonian $\tilde{H}$ in the configuration space/many-body state basis: ${\rm IPR}_\alpha = \sum_{i=1}^{\mathcal N} |\langle i | \psi_\alpha \rangle|^4$, where $\ket{i}$ is an element of the many-body state in $S_z$ basis, indexed by $i=1, \ldots, \mathcal N$, and $\mathcal N$ is the size of the Hilbert space. The volume occupied by an ergodic state is proportional to $\mathcal N$, while a non-ergodic state occupies a vanishing fraction $\ll \mathcal N$. Thus, the towers of large values of IPR seen for certain eigen-energies (highlighted in red) are a signature of non-ergodicity.  The parent homogeneous Heisenberg Hamiltonian (Fig.~\ref{fig:2}(b)) does not show such towers.  Similar conclusions can be also be obtained from the entanglement entropy as shown in the Appendix \ref{appendix_B}.

Figure \ref{fig:2}(c) further demonstrates the characteristic dynamics with revivals associated with QMBS \cite{scar_states_Turner2018}. 
For a quench from an initial state having a sufficiently large overlap with the QMBS eigenstates of the target Hamiltonian (a random superposition of eigenstates whose $\ln {\rm IPR}$ is normally distributed around a large $\ln {\rm IPR}_0 =-3.0$ with a small standard deviation $\sigma_{\ln {\rm IPR}} = 0.5$), periodic revivals appear clearly with the scar Hamiltonian. However, the same initial state shows a fast decay of the return probability (initial state overlap) in the case of the homogeneous Heisenberg model.  In this example, we demonstrated that EHC can be used to discover new Hamiltonians with weak non-ergodic properties distinct from previously known models for scarring.  The QMC allows us to reliably reach large system sizes of up to $N=256$.  

\section{Many-Body Localization}

\begin{figure*}[t]
\centerline{
\includegraphics[width=1.0\linewidth]{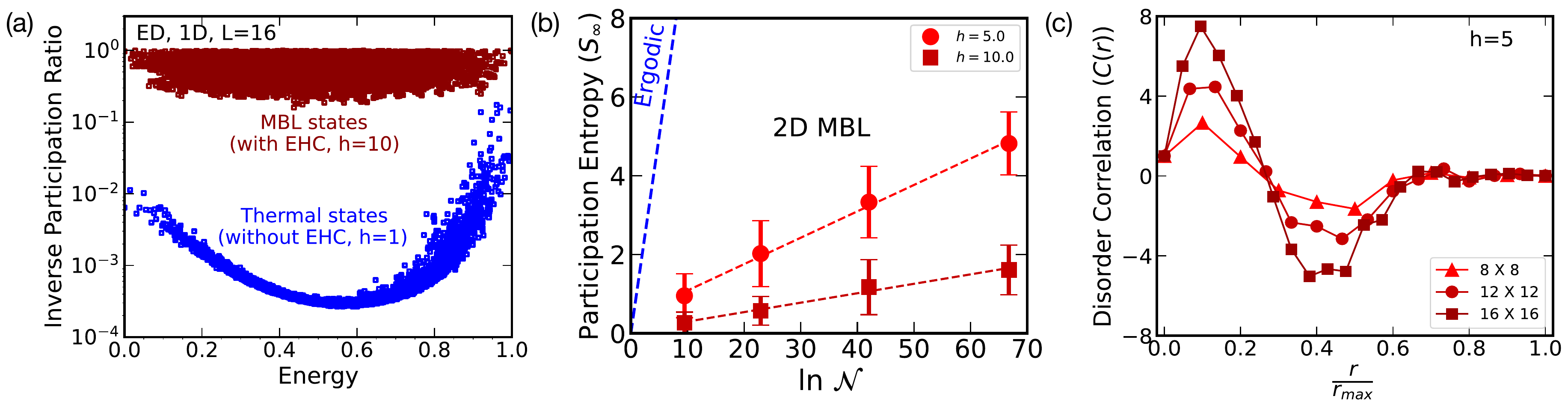}
}
\caption{
{\bf Characteristics of the MBL model:}
(a)~Comparison of the non-ergodic properties of all eigenstates with exact diagonalization of our EHC built MBL Hamiltonian (red) and the standard disordered Heisenberg model without EHC (blue) using the inverse participation ratio.  The EHC built Hamiltonian exhibits eigenstates with much larger IPR values than any typical thermal states demonstrating stronger non-ergodicity behaviour.  (b)~Scaling of disorder-averaged participation entropy, $S_\infty$ of the state $|\Psi \rangle$ with the Hilbert space size, ${\cal N}$ computed with the EHC-QMC method in 2D for disorder values $h=5.0$ and $10.0$. As discussed in the text, the slope indicates multifractal scaling and non-ergodic properties. (c)~Behavior of the disorder-averaged correlation function $C(r)$ of the new disorder in 2D obtained with the EHC-QMC method (see text). $C(r)$ is shown at fixed disorder ($h=5.0$) for three different lattice sizes. We use 1280 disorder realisations for averaging the observables in (b) and (c)} 
\label{fig:3}
\end{figure*}

For MBL Hamiltonians all excited states in a finite energy range exhibit nonergodic properties. First raised in the context of the (vanishing) conductivity of disordered wires~\cite{Gornyi2005}, these systems, where disorder plays a crucial role, are the subject of intense theoretical and experimental interest. While the experimental signatures of MBL appear identical in one and two dimensions~\cite{schreiber2015,Smith2016,choi2016,Kondov2015}, the very existence of MBL remains theoretically controversial due to an instability mechanism called the ``thermal avalanche''~\cite{DeRoeck2017,Potirniche2019} that makes MBL impossible in two and higher dimensions, and possibly pushing the critical disorder strength in one dimension to large or even infinite values~\cite{Suntajs2020,Sierant2020,mbl_infinite_no3,mbl_infinite_no4,Sels2021_MBL_hc_inf,Morningstar2022}.

We note that these works assume uncorrelated disorder.  Including disorder correlations can significantly change these conclusions.  For example, in a recent work \cite{Darryl2022}, we have shown that confining potentials that are ubiquitous in experimental realizations of MBL (and a form of correlated disorder) can shift the critical dimension of MBL from d=1 to d=2.  Similarly, Stark MBL has also been shown to arise in the absence of disorder~\cite{Stark-MBL,starkmbl2,Stark-MBL_Doggen,starkmbl3}, and quasi-periodic potentials also show signatures of MBL properties~\cite{quasimbl1,quasimbl3,quasimbl4,quasimbl5}. These potentials may generically be thought of as the correlated part of an otherwise uncorrelated disorder potential, if any.  Below we will find that disorder correlations emerge from our EHC procedure even when we start with a parent Hamiltonian with uncorrelated disorder.  Similar correlations were observed by Ref.~\cite{Dupont2019a} when using EHC in conjunction with DMRG to investigate MBL in d=1.  Here we apply the EHC to study MBL in 2D which is made feasible by the QMC.

We consider the spin-$1/2$ Heisenberg model in a random, but possibly correlated field, which acts as quenched disorder. In two dimensions, it is described by the Hamiltonian
\begin{align}
\tilde{H} = \tilde{J}\sum_{\langle i,j\rangle} {\bf S}_i \cdot {\bf S}_j + \sum_{i} \tilde{h}_i S_i^z,
\label{eq1}
\end{align}
\noindent where $\tilde{J}$ is the coupling strength which we set to unity, ${\bf S}_i$ is the spin operator at site $i$, 
$\langle i,j\rangle$ indicates nearest-neighbour sites, and $\tilde{h}_i$ is the local disordered magnetic field. Our input state is the ground state of $H$ with uncorrelated disorder  $h_i\in [-h,h]$.  Beyond a critical disorder strength $h > h_c \approx 2.35$, $H$ has a Bose glass ground state  (see e.g. Ref.~\cite{SF_BG_2D_HCB} and Appendix~\ref{appendix_F}).  The Bose glass ground state is insulating and characterized by correlators $\langle {\bf S}_i \cdot {\bf S}_j \rangle$ that decay exponentially with distance~\cite{SF_BG_2D_HCB}.  This is different from the clean and homogeneous Heisenberg Hamiltonian considered in the previous section whose ground state has quasi long-range antiferromagnetic order~\cite{Affleck_1998}.  As with the QMBS Hamiltonian, the Hilbert space size of the zero net spin sector is $\frac{N!}{((N/2)!)^2}\sim N^{-1/2}2^N$, growing exponentially with system size. 

We proceed to construct the target Hamiltonian hosting the ground state of $H$ as an excited state. We select our EHC basis operators as ${\mathcal O}_0 = \sum_{\langle i,j\rangle} {\bf S}_i \cdot {\bf S}_j$ and ${\mathcal O}_i = S_i^z, \;\; i=1,\ldots,N$ so that the number of operators $N_{\mathcal O}=N+1$ and the only possible outcome of EHC is to change the disorder configuration.  We calculate the $(N+1)\times (N+1)$ covariance matrix ${\mathcal C}$ using QMC.  Details of the calculation and calibration checks performed with ED is discussed in the Appendix~\ref{appendix_A}. As described earlier, we look for (near) zero eigenvalues of ${\mathcal C}$. Two of these are trivially zero, corresponding to the original Hamiltonian and the total spin operator, $S^z=\sum_i S^z_i$, respectively. So in practice, we look for the third smallest eigenvalue to get our target Hamiltonian $\tilde{H}$.   This eigenvalue scales to zero in the thermodynamic limit (see Sect.~\ref{sec5}).

Figure \ref{fig:3} illustrates the main properties of the obtained target Hamiltonian that is consistent with many-body localization.  We plot the IPR of all the eigenstates (similar to Fig.~\ref{fig:2} (a),(b)).   In dark-red squares, we show the target Hamiltonian $\tilde{H}$ obtained from a 1D version of $H$ at $h=10$, and in blue squares, the parent Hamiltonian $H$ at $h=1$ (see \eqref{eq1}).  For this illustration, we consider the 1D case because we are restricted to ED for this characterization of the spectrum, and the ED has too strong finite-size effects in 2D. We note that the parent Hamiltonian $H$ is the paradigmatic model of MBL, which, at $h=1$, is in the ETH phase where all states have ergodic/thermal properties.   This is also manifest in the low values of IPR proportional to the inverse of the Hilbert space size $\mathcal{N}$. In contrast, the eigenstates of the target Hamiltonian $\tilde{H}$ have large values of the IPR -- orders of magnitude larger than those of ergodic states for the same Hilbert space size. This is a clear signature of the MBL nature of the target Hamiltonian, albeit in 1D and for small system sizes.

Next we look at the properties in 2D using QMC that enables us to describe the properties of $\ket{\Psi}$ up to very large Hilbert space sizes $\sim \mathcal{O}(2^{250})$. In Fig.~\ref{fig:3} (b), we represent the participation entropy $S_{\infty}=\lim_{q\to\infty}\frac{1}{1-q}\ln\left(\sum_{i=1}^{\mathcal N}| \langle \Psi|i \rangle|^{2q}\right)$ of the state $\ket{\Psi}$, where $\ket{i}$ are many-body states in $S_z$ basis, as a function of the size of the Hilbert space $\mathcal N$. The dashed lines show the scaling of $S_{\infty}$ for $q\rightarrow \infty$ with ${\cal N}$ where $S_{\infty} = D \ln{\cal N} + c$.  $D$ is the multifractal dimension and $c$ is a constant.  The benchmark (shown in blue) is ETH ergodic regime where $D = 1$.   The data for our target Hamiltonians are obtained for very large 2D samples (up to $10 \times 10$) using QMC and all have $D<1$.  This indicates a vanishing fraction of the states in configuration space contribute to $\ket{\Psi}$ demonstrating its  non-ergodic properties \cite{multifractal_scaling_Laflorencie}.  Additional characterization of the MBL properties of the target Hamiltonian is shown in Section~\ref{appendix_B}.

Finally, in Fig.~\ref{fig:3} (c), we show that, despite $H$ having uncorrelated disorder, the new Hamiltonian $\tilde{H}$ always has correlated disorder. The new disorder can be expressed as $\tilde{h}_i = h_i + \Delta h_i$, with $\Delta h_i$ used to define a correlation function $C(r) = ( \sum_{d_{ij}=r} \Delta h_i \Delta h_j ) /( \sum_i \Delta h_i^2 )$, where $d_{ij}$ is the distance between sites $i$ and $j$. This correlation function is plotted as a function of $r/r_{\rm max}$, where $r_{\rm max}$ is the maximum distance in our 2D sample. We further show how the emergent correlations scale with system size. We clearly observe self-similar disorder correlations as the scaled distance between sites increases, with one prominent maximum and minimum. The strength of the correlation, and thus of the square of the confining potential, scales with the linear dimension $L$ of the system ($N=L^2$), mirroring previous findings~\cite{Darryl2022} about MBL stabilized by a confining potential in two dimensions.
 
The EHC-QMC method allows us to obtain new Hamiltonians with MBL properties in two dimensions which was thought to be impossible due to thermal avalanches. The difference, of course, is the strong disorder correlations that emerge from the method.  Starting from the localized ground state of a Hamiltonian with weak uncorrelated disorder, we obtain a target Hamiltonian with strong correlated disorder whose entire spectrum is non-ergodic.  In addition, Hamiltonians corresponding to higher eigenvalues of the covariance matrix also exhibited MBL properties (not shown).

\begin{figure*}[ht!]
\centerline{
\includegraphics[width=0.95\linewidth]{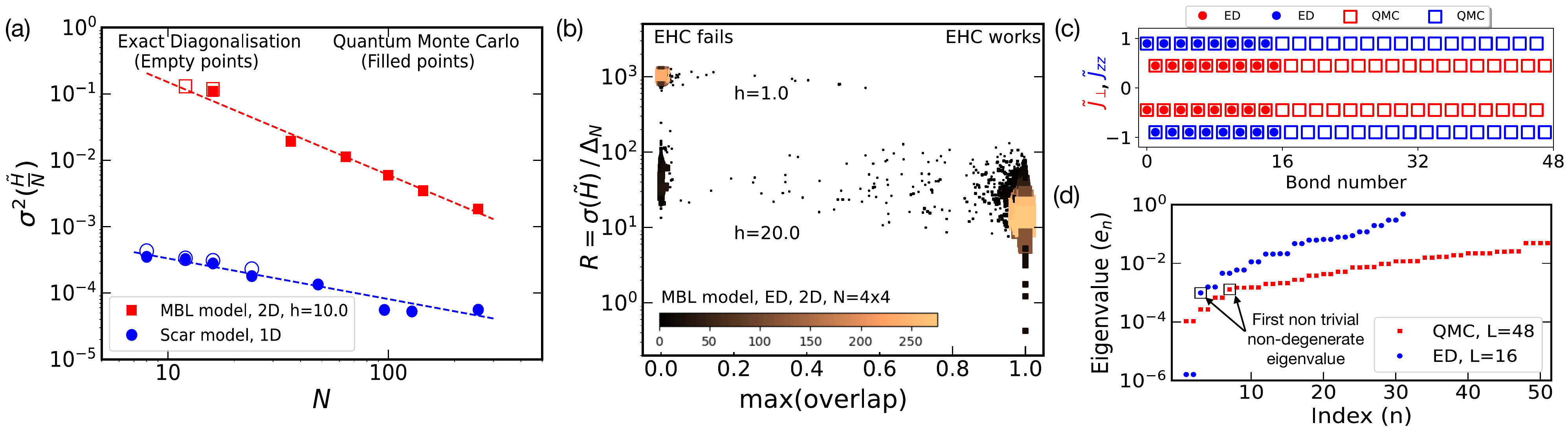}
}
\caption{ \textbf{Reliability of the method:} (a) Algebraic decrease of energy density variance with system size for both the many body scar and MBL Hamiltonians. Small system sizes are studied with ED and large system sizes using QMC.  The red dashed line is a power-law fit, $\sigma^2(\tilde{H}/N) \approx 3.7N^{-1.4}$ to the numerical data of our MBL model averaged over 1280 disorder realisations.  The blue dashed line is a power-law fit, $\sigma^2(\tilde{H}/N) \approx 0.01N^{-0.6}$ to the numerical data of our scar model.
(b) 2D histogram with 2000 disorder realizations (color indicating the frequency) comparing the relative residue, $R = \sigma(\tilde{H})/\Delta_N$ expressed as the ratio of the energy standard deviation (the square root of the energy variance considered in panel (a)) to the mean level spacing $\Delta_N$ (see Appendix~\ref{appendix_C} for details), calculable using QMC, with the maximal overlap (defined in the main text and only obtainable in ED). Small values of $R$ or max(overlap) values close to 1 indicate that EHC procedure works. EHC does not work reliably at weak disorder ($h=1$ is shown) where most disorder configurations have large $R$ and small max(overlap) values.  By contrast, EHC works for most disorder configurations at large disorder (shown for $h=20$).  (c)-(d) Comparison between QMBS results for large system size ($L=48$), obtained from EHC-QMC,  and small system size ($L=16$), obtained from ED. The coupling parameters $\tilde{J}_\perp$ and $\tilde{J}_{zz}$ defining the target Hamiltonian obtained by EHC-QMC agree perfectly with ED. This is a non-trivial observation since both use different eigenvalues of the covariance matrix as shown in (d). 
} 
\label{fig:4}
\end{figure*}

\section{Discussions}
\label{sec5}
\subsection{Reliability of the Method}

Since the EHC is an approximate method, it is necessary to assess its reliability. The eigenvalue $e_n$ of the covariance matrix determines the variance of the target Hamiltonain $\tilde{H}$ in the state $\ket{\Psi}$.  We have $\sigma^2(\tilde{H})\equiv \langle \Psi \vert \tilde{H}^2 \vert \Psi \rangle - \langle \Psi \vert \tilde{H} \vert \Psi \rangle^2$, where $\sigma^2(\tilde{H}) = e_n/(\tilde{J_{\perp}}^2)$ for the QMBS case, and  $\sigma^2(\tilde{H}) = e_n/(\tilde{J}^2)$ for the MBL case.  Here $n$ is the index of the considered eigenvalue.  In Fig.~\ref{fig:4}(a), we display the energy density variance $\sigma^2(\tilde{H}/N)$ as a function of system size for both cases, where the data is obtained using ED at small system sizes and using QMC at large sizes. The MBL data is disordered averaged. We clearly observe a power-law decay and very good agreement between ED and QMC data. Since a vanishing energy variance signals that $\ket{\Psi}$ is an exact eigenstate of $\tilde{H}$, this decrease and consistency between ED and QMC is a first indication that our EHC approach works reliably even at large system sizes.

\begin{figure*}[t]
\centerline{
\includegraphics[width=0.8\linewidth]{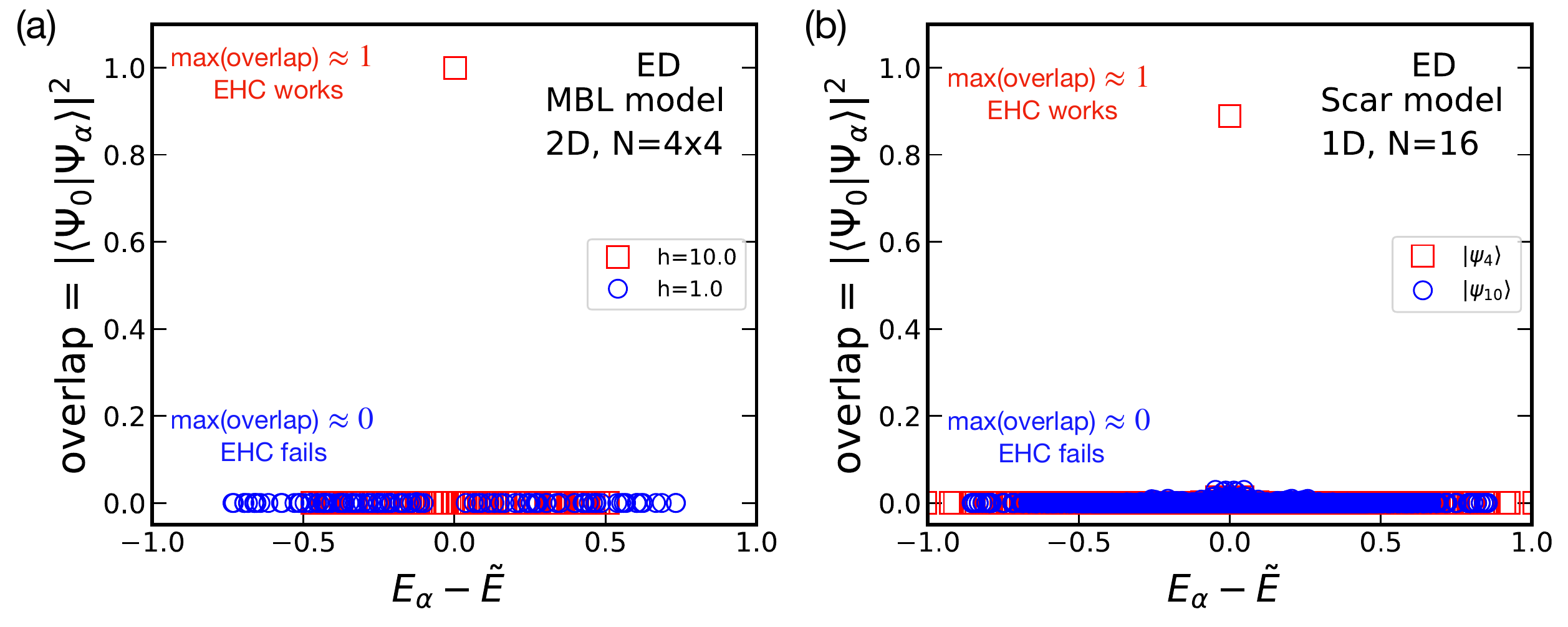}
}
\caption{(a)~We compute the overlap of the ground state of our MBL model in 2D, $| \Psi_0 \rangle$, of the parent Hamiltonian, (\eqref{SM_eq1}) and the eigenstates of the EHC built Hamiltonian, $\tilde{H}$, close to the energy $\tilde{E} = \langle \Psi_0 | \tilde{H} | \Psi_0  \rangle$, for a given disorder configuration with different disorder strengths. 
We see that for strong disorder ($h=10$), the overlap is maximum ($ \approx 1$) for a single eigenstate closest to $\tilde{E}$, indicating that EHC has successfully discovered a $\tilde{H}$ hosting $| \Psi_0 \rangle $ as an exact eigenstate. In contrast, for weak disorder ($h=1$), the overlap $\ll 1$, indicating the mapping to a superposition of eigenstates of $\tilde{H}$ and a failure of the EHC.
(b)~We compute the overlap of the ground state of our 1D model \eqref{SM_eq2}, $| \Psi_0 \rangle$, and the eigenstates of EHC built Hamiltonian, $\tilde{H}$, for two different input eigenvectors of the covariance matrix. We see that the ground state has maximum overlap with only a single eigenstate of the EHC built model, when the input is the $4^{th}$ eigenvector (that hosts couplings of the scar model). On the other hand, when the input eigenvector is a different one (say $10^{th}$ eigenvector), we observe that the ground state shows vanishing overlaps with all the eigenstate of the EHC built model, an indication of the failure of the EHC mapping.} 
\label{fig_SM4}
\end{figure*}

However, this does not guarantee that $\ket{\Psi}$ maps to a single eigenstate of the new Hamiltonian. In fact, as the excited states of a many-body system have an exponentially large density, $\ket{\Psi}$ could correspond to a superposition of eigenstates. This limitation is common to all such approximate methods~\cite{Khemani2016,Yu2017,Wahl2019}.
To address this question, we use exact diagonalization, keeping in mind the limited applicability to small system sizes.  We determine the eigenstates $| \psi_{\alpha} \rangle$  of $\tilde{H}$ close in energy to that of $\ket{\Psi}$,  and calculate their corresponding overlap $O_{\alpha} = |\langle \Psi | \psi_{\alpha} \rangle|^2$. In particular, we focus the maximum value of these overlaps, max(overlap)$\equiv {\rm max}_\alpha O_{\alpha}$. Values of max(overlap) close to 1 indicate that EHC works i.e. $\ket{\Psi}$ maps mainly to a single eigenstate of $\tilde{H}$. 

An alternate figure of merit that is accessible to QMC is the relative residue $R = \sigma(\tilde{H})/\Delta_N$, where $\Delta_N$ the mean level-spacing of the many-body target Hamiltonian $\tilde{H}$.  Fig.~\ref{fig:4}(b) compares $R$ with max(overlap) for the MBL case.  We present a 2D histogram of the $R$ and max(overlap) values for over $2000$ disorder realizations (with the color indicating the frequency).  We clearly observe that the EHC does not work at weak disorder (e.g. $h=1$) where most disorder configurations have large $R$ and small max(overlap). By contrast, EHC works for most (but not all) disorder configurations at large disorder ($h=20$) despite all disorder configurations having similar $R$ values.  It is not surprising that the EHC approach always works when $R < 1$, i.e. when the error in the energy is small compared to $\Delta_N$ (for $R < 1$, we always find $\text {max(overlap)} \approx 1$).  However, it might be counter intuitive that EHC also works at large disorder when $R > 1$.  This is because even if the energy resolution is not sufficient (as is the case in QMC), the non-ergodic properties associated with MBL nevertheless allow EHC to work (see Section~\ref{appendix_E}).  

For the QMBS case, we have also verified that system sizes up to $L=16$ have large values of max(overlap)$\ge 0.8$ indicating that the EHC method works (see Section~\ref{appendix_E}).  At even larger system sizes -- only accessible through QMC -- such a characterization is not possible.  Instead, we compare directly the QMC-obtained target Hamiltonian at large sizes with the ED results at small sizes.  In Fig.~\ref{fig:4}(c), we show that the coupling parameters $\tilde{J}_\perp$ and $\tilde{J}_{zz}$, which define the target Hamiltonian $\tilde{H}$, obtained by EHC-QMC align perfectly with those derived from ED. This is significant since the considered eigenvalue of the covariance matrix for QMC corresponding to the lowest non-trivial, non-degenerate eigenvalue, is the 8th eigenvalue for $L=48$, whereas it is the 4th eigenvalue for ED.  This is shown in Fig.~\ref{fig:4}(d).  We interpret this perfect correspondence as further evidence of the reliability of the EHC-QMC approach.

\subsection{Mapping to excited states of \\ the target Hamiltonians}
\label{appendix_E}

The accuracy of the EHC mapping is inferred from a decaying behavior of the variance of the energy density with increased system size, and thus vanishing in the thermodynamic limit. But due to the exponentially large degeneracy of excited-eigenstates close to energy, $\tilde{E}$, the question remains, whether $\Psi_0$ maps to a single eigenstate of $\tilde{H}$ or a superposition of eigenstates. To address this, we perform Exact Diagonalization (ED) calculations using the state-of-the-art {\it Quspin} package\cite{Quspin_2017} and compute the overlap of the ground state of the parent Hamiltonian, $H$, and the eigenstates of the EHC built Hamiltonian, $\tilde{H}$.

In Fig.~\ref{fig_SM4}(a), we show the overlap, $O_{\alpha} = |\langle \Psi_{0}| \Psi_{\alpha} \rangle|^2 $ of the actual ground state, $\Psi_{0}$ and the eigenstates of the target Hamiltonian $\tilde{H}$, close to the energy, $\tilde{E} = \langle \Psi_{0} | \tilde{H} | \Psi_{0} \rangle$, for a single disorder realisation of strong and weak disorder values on a 2D lattice of sixe $4 \times 4$. We find that in the strong disorder case, the overlap is maximum $O_m \approx 1$ for a single eigenstate and vanishing for the remaining eigenstates. This indicates the EHC mapping to only one eigenstate in the strong disorder limit. On the other hand for the weak disorder configuration, the overlap is finite for several eigenstates. This is indicative of the fact that the ground state maps to a superposition of eigenstates.

In a similar fashion, in Fig.~\ref{fig_SM4}(b), we show the overlap, $O_{\alpha} = |\langle \Psi_{0}| \Psi_{\alpha} \rangle|^2 $ of the ground state, $\Psi_{0}$ and the eigenstates of the target Hamiltonian $\tilde{H}$, for two different input eigenvectors of the covariance matrix (say $4^{th}$ and $10^{th}$ eigenvectors) of our scar model in 1D \eqref{SM_eq2}. We see that the ground state has maximum overlap with only a single eigenstate of the EHC built model, when the input eigenvector is the $| \Psi_4 \rangle$, the one hosting the couplings of our scar model. On the other hand, when the input eigenvector is a different one (say $| \Psi_{10} \rangle$), we observe that the ground state shows vanishing overlaps with all the eigenstate of the EHC built model, an indication of the failure of the EHC mapping.
\vspace{0.2in}
\begin{figure*}[t]
\centerline{
\includegraphics[width=0.8\textwidth]{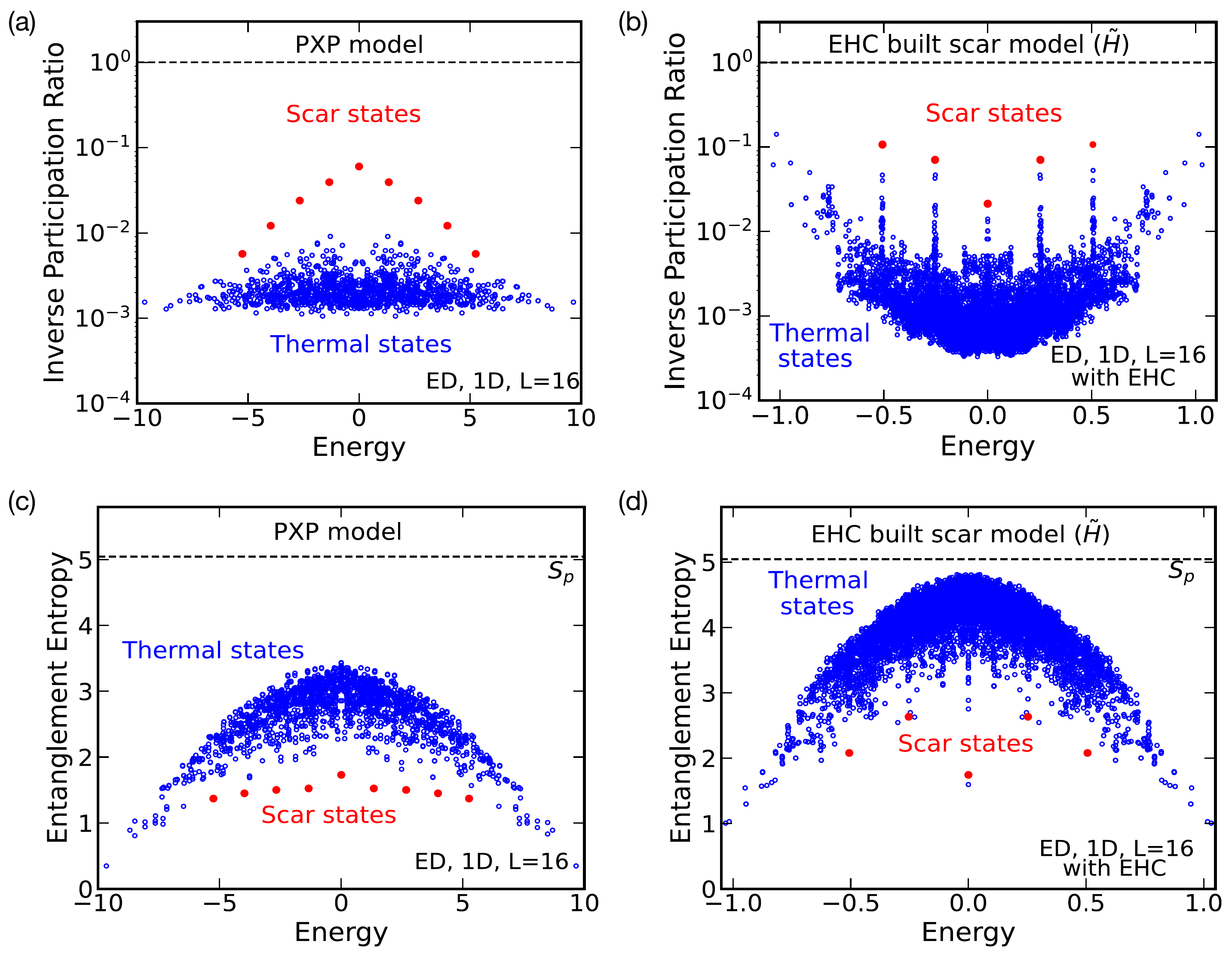}
}
\caption{(a-b)~We show the behavior of the inverse participation ratio (IPR) of the eigenstates of the PXP model and our EHC built scar model on 1D chain of $L=16$. It can be clearly seen the IPR values of the scar eigenstates are clear outliers from the typical thermal eigenstates.
(c-d)~We show the behavior of half-chain entanglement entropy of the eigenstates of the PXP model and our EHC built scar model on a chain of $L=16$. It can be seen that the scar states (highlighet in blue) have lower entanglement entropy than the thermal states in the same energy density. }
\label{fig_SM2}
\end{figure*}

\subsection{Application in known models}
\label{appendix_B}
We show the behavior of relevant physical observables obtained with our method and compare them with previously known models exhibiting scar and MBL properties respectively. 

A key observable to study the non-ergodic properties of the eigenstates is the inverse participation ratio(IPR). For a single particle eigenstates, $IPR \sim 1.0$ indicates fully localised nature of the state, whereas  $IPR \sim 1/N_H$, where $N_H$ is the Hilbert space size, indicates a fully delocalised nature. For many-body states this behavior is not straightforward.  
We perform Exact Diagonalization (ED) calculations in a 1D chain with periodic boundary, using the state-of-the-art {\it Quspin} package\cite{Quspin_2017} to access all the eigenstates of our many-body Hamiltonians.

\begin{figure}[t]
\centerline{
\includegraphics[width=0.4\textwidth]{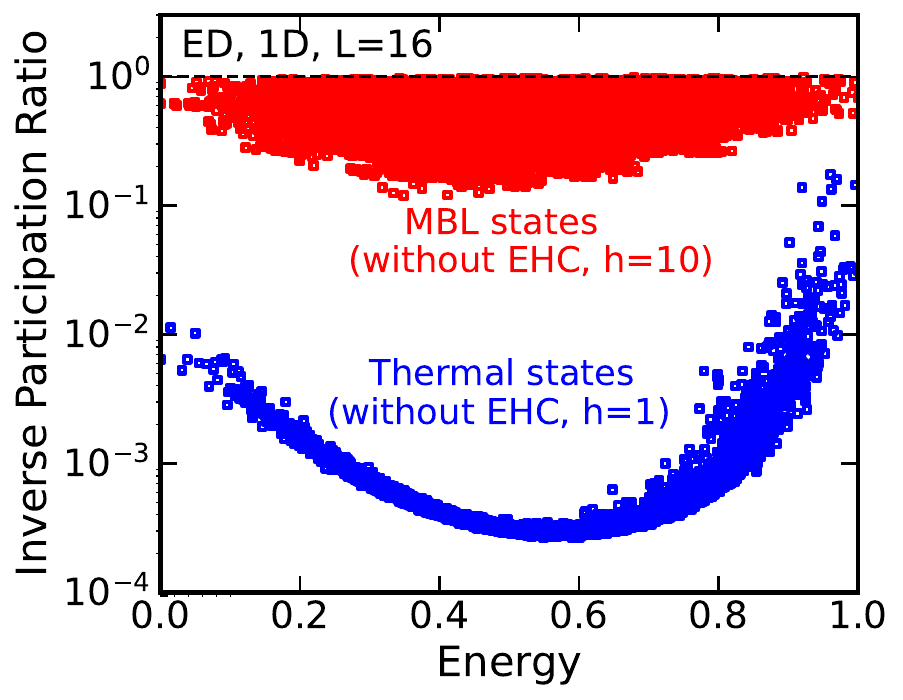}
}
\caption{We show the behavior of the inverse participation ratio (IPR) of the eigenstates of the disordered Heisenberg model for a typical disorder realisation for weak ($h=1$) and strong ($h=10$) disorder values. It can be clearly seen the IPR values of the eigenstates in the MBL regime are much higher than those in the ETH regime. }
\label{fig_SM1}
\end{figure}

{\it Scar model :}
A well known model describing the physics of Rydberg atom chain and hosting scar states is the PXP model~\cite{PXP_2008}. We compare the properties of our EHC built scar model with that of the PXP model.

In fig.\ref{fig_SM2}, we compare the inverse participation ratio(IPR) of the eigenstates of our EHC built scar model, ${\tilde{H}}$ and that of the PXP model as a function of energy. As expected, we see that majority of the eigenstates have vanishing IPR values, typical of any thermal eigenstate. In addition, we clearly observe several high-energy excited states exhibiting markedly-higher IPR than the typical thermal eigenstates at given energies. These are the quantum many-body scar states. It should be noted that the scar-states in the PXP model appear in the eigen-spectrum with equally spaced energies. However, this is not the case for the scar states in our EHC built model. 

A key characteristic of scar states is the 
visibly lower value of entanglement entropy in the bipartite entanglement entropy spectrum.
In Fig.~\ref{fig_SM2}, we compare the bipartite entanglement entropy spectrum of our EHC built scar model ${\tilde{H}}$, and the PXP model as a function of the energy. Like the PXP model, our EHC built model hosts several high-energy excited states exhibiting markedly lower entanglement entropy than the typical thermal eigenstates at given energies, identified as scar states. Furthermore, these scar-states appear over the entire energy band.

{\it MBL model :}
With our EHC approach, we build Hamiltonians with correlated disorder that exhibit non-ergodic properties in the excited states. We compare the IPR behavior of all eigenstates of our EHC built target Hamiltonian, ${\tilde{H}}$, vis-a-vis the prototypical MBL model, that is the disordered Heisenberg model, $H$, \eqref{SM_eq1}. We use disorder values $h=1.0$ and $10.0$ for studying the ergodic and the localised regimes respectively. 
In figure \ref{fig_SM1}, we show the IPR of all the eigenstates of the disordered Heisenberg model, \eqref{SM_eq1}, as a function of the energy.

\vspace{0.2in}
\section{Conclusion and Perspectives} 


We have extended the scope of the EHC by using QMC simulation of ground states to systematically build quantum many-body Hamiltonians with non-ergodic excited states on large systems. Systems exhibiting non-ergodic properties have garnered significant interest recently because they can evade standard thermalization, enabling the preservation of quantum information over extended or even infinite times, even at infinite temperature. Previously, such systems were typically discovered through serendipity or arguments based on elaborate intuition and symmetry.

Our EHC-QMC approach takes a different path by systematically constructing such Hamiltonians through a procedure where the specific symmetries or correlations of the parameters defining the system emerge naturally. The key idea is to start from ground states that inherently possess non-ergodic properties and then search for target Hamiltonians that host such ground states as highly excited eigenstates.  The EHC construction engineers the Hamiltonian supporting non-ergodic excited states, while the QMC provides an accurate description of ground state properties harnessing its capacity for large system sizes in higher dimensions.  Our two examples illustrate that our approach provides a useful method to systematically construct Hamiltonians with non-ergodic states.  While our focus here has been on the non-ergodic properties of highly excited states, our method allows us to systematically construct new Hamiltonians with desired excited state properties inherited from a specific ground state. 

In condensed matter systems, ground states are known to exhibit unique and interesting properties, such as the quasi-long-range ordering seen in the QMBS case we have described, but also superconductivity, Wigner crystallization, the quantum Hall effect, and topological ordering, among others. With our EHC-QMC method, these ground state properties can be promoted to highly excited states.  This opens up the possibility for symmetry breaking and long-range order to occur even at infinite temperature.  For example, we speculate that this algorithm could engineer a system with an ``infinite temperature" superconducting state, as provocative as that prospect may seem.

\section*{Data Availability}
The data that support the findings of this article are openly
available at https://doi.org/10.6084/m9.figshare.28040900

\section*{Acknowledgements} 
We thank Nicolas Laflorencie, Maxime Dupont, Jeanne Colbois, and Rubem Mondaini for helpful discussions, and Miguel Dias Costa for assistance with the parallelization of our code. This work is supported by the Singapore Ministry of Education AcRF Tier 2 Grants No. MOE2017-T2-1-130 and No. MOE-T2EP50222-0005, and AcRF Tier 3 Grant No. MOE-MOET32023-0003, and made possible by allocation of computational resources at the Centre for Advanced 2D Materials (CA2DM), and the Singapore National Super Computing Centre (NSCC). FFA thanks support from the W\"urzburg-Dresden Cluster of Excellence on Complexity and Topology in Quantum Matter ct.qmat (EXC 2147, project-id 390858490). GL acknowledges the support of the projects GLADYS ANR-19-CE30-0013 and MANYLOK ANR-18-CE30-0017 of the French National Research Agency (ANR), and by the Singapore Ministry of Education Academic Research Fund Tier I (WBS No. R-144-000-437-114).

\section*{Author contributions statement}
FFA, GL, PS, and SA conceived the project.  HKT and BJJK did the initial proof of concept under the supervision of FFA, PS, and SA. NS, HKT and DCWF developed the codes and implemented the research under the guidance of PS, GL, and SA.  NS, HKT, DCWF and GL analyzed the results. All authors discussed the results. NS, DCWF, GL, PS, and SA wrote the paper.


\appendix

\section{Computation of the Covariance Matrix with the Stochastic Series Expansion (SSE) QMC}
\label{appendix_A}

The eigenstate-to-Hamiltonian construction (EHC) approach crucially needs the evaluation of a quantum covariance matrix, which mainly involves the computation of a collection of expectation values of correlators with respect to the ground state wavefunction. 
In SSE QMC~\cite{Sandvik1997,Sandvik1999}, ground state expectation values for finite size systems is obtained by choosing a sufficiently large inverse temperature $\beta$ (that depends on the system size). The spectrum of any finite-size system is discrete and for simulations performed at temperatures smaller than the finite-size gap (between the ground state and the first excited state), contributions from  higher energy states  are exponentially suppressed, yielding ground state expectation values for the finite size system. Estimates for thermodynamic quantities are then obtained through a simultaneous finite-size and finite-temperature scaling (the temperature for each simulation is adjusted carefully to ensure that it is smaller than the finite size gap). In the literature~\cite{SSE_rev_Sandvik}, this approach has been successfully applied by all finite-temperature QMC algorithms (SSE, determinant QMC, world line QMC, path integral QMC) to investigate the ground state phases of interacting spins, bosons and fermions, both with and without disorder. We perform our SSE-QMC simulations at low finite temperature, with $\beta$ values changing from $\beta = 2L$ to $\beta = 4L$ and further to $\beta = 8L$, where L is the linear dimension of our system size under study. We observe that the ground state energy converges readily with increasing $\beta$, and hence we set $\beta = 8L$ to ensure we are in the ground state for computation of various observable of interest.

Having obtained the ground state (say $\ket{\psi}$) of a parent Hamiltonian (say $H$), we choose a set of local operators $\{\mathcal O_i\}_{i=1,N_\mathcal O}$, as our EHC basis operators, such that $H=\sum_{i=1}^{N_\mathcal O} h_i \mathcal O_i$. We emphasize that the choice of this set of local operators is not unique, and mainly dictated by the problem at hand and the ease of numerical evaluation.

The next step is to compute the covariance matrix $\mathcal C$ whose elements are given by:
\begin{equation}
\mathcal C_{ij}=\bra{\psi} \mathcal O_i \mathcal O_j\ket{\psi}-\bra{\psi} \mathcal O_i\ket{\psi}\bra{\psi} \mathcal O_j\ket{\psi} ; .
\end{equation}

It should be noted that unlike the Hamiltonian, whose dimension is exponential in system size, the dimension of a well chosen covariance matrix is linear in system size. Thus the covariance matrix can be diagonalized numerically by well established computational packages, and its eigenvalues and eigenvectors can be obtained precisely.  

An eigenvector of the covariance matrix $\mathcal C$ provides the coefficients $\tilde{h}_i$ defining a target Hamiltonian $\tilde{H}=\sum_{i=1}^{N_\mathcal O} \tilde{h}_i \mathcal O_i$. Its associated eigenvalue represents the energy variance of $\ket{\psi}$ with respect to the target Hamiltonian $\tilde{H}$. That is, if this eigenvalue is zero, then $\ket{\psi}$ is an exact eigenstate of $\tilde{H}$. On the other hand, if this eigenvalue is non-zero (which typically happend on finite system sizes), then it's scaling behavior with increased system size needs to be studied. If the eigenvalue decays at least as a power-law fashion, then it is argued that $\ket{\psi}$ is an approximate eigenstate of $\tilde H$ \cite{Dupont2019a,Dupont2019b}.

\subsection{Scar model}

The Hamiltonian of interest is the $S=1/2$ antiferromagnetic Heisenberg model on a 1D chain with periodic boundary,
\begin{align}
H= \sum_{i} [ (J/2) (S_i^{+}  S_{i+1}^{-} + S_i^{-}  S_{i+1}^{+}) + J S_i^z S_{i+1}^z ],
\label{SM_eq2}
\end{align}

We write the Hamiltonian as above to distinguish between its diagonal part ($S_i^z S_{i+1}^z$) and off-diagonal part ($S_i^{+} S_{i+1}^{-} + S_i^{-} S_{i+1}^{+}$) on each bond in the standard basis of diagonal z spin components. In SSE QMC, we use the Taylor expansion to expand the exponential part of the partition function. Hence, the partition function can be written as a sum of the products (strings) of different Hamiltonian operators with the inverse temperature $\beta$ as its order, in which its sequence is usually referred to as an operator string (see chapter-5 of Ref~\cite{SSE_rev_Sandvik} for further details).

We choose $2N$ EHC basis operators to construct the covariance matrix, ${\mathcal C}$, such that ${\mathcal O}_i= (1/2)(S_i^{+} S_{i+1}^{-} + S_i^{-} S_{i+1}^{+})$ and ${\mathcal O}_{i+N}=S_i^z S_{i+1}^z$  with $i=1,\ldots,N$. Thus the dimension of the covariance matrix in this case is $2N \times 2N$.

As the parent Hamiltonian follows translation invariance, the covariance matrix constructed as above also requires to follow translation invariance. Thus the correlation, $\langle {\mathcal O}_i {\mathcal O}_j \rangle$ should depend only on $i-j$. With QMC we compute,
\begin{align}
\langle {\mathcal O}_i {\mathcal O}_j \rangle =
\frac{1}{(-\beta^2)} \langle (n-1) N(i,j) \rangle
\end{align}
where, $N(i,j)$ denotes the number of times the indices $i$, $j$ appear next to each other in the operator string\cite{Sandvik1992}. QMC provides statistical estimations of these correlators and thus may not fulfill this invariance exactly. Therefore, we impose it such that $\langle {\mathcal O}_i {\mathcal O}_{i+r} \rangle \equiv \sum_i \langle {\mathcal O}_i {\mathcal O}_{i+r} \rangle / N $, during the evaluation of the matrix elements of the covariance matrix.

\begin{figure*}[t]
\setcounter{figure}{0}
\renewcommand{\thefigure}{A\arabic{figure}}
\centerline{
\includegraphics[width=0.99\textwidth]{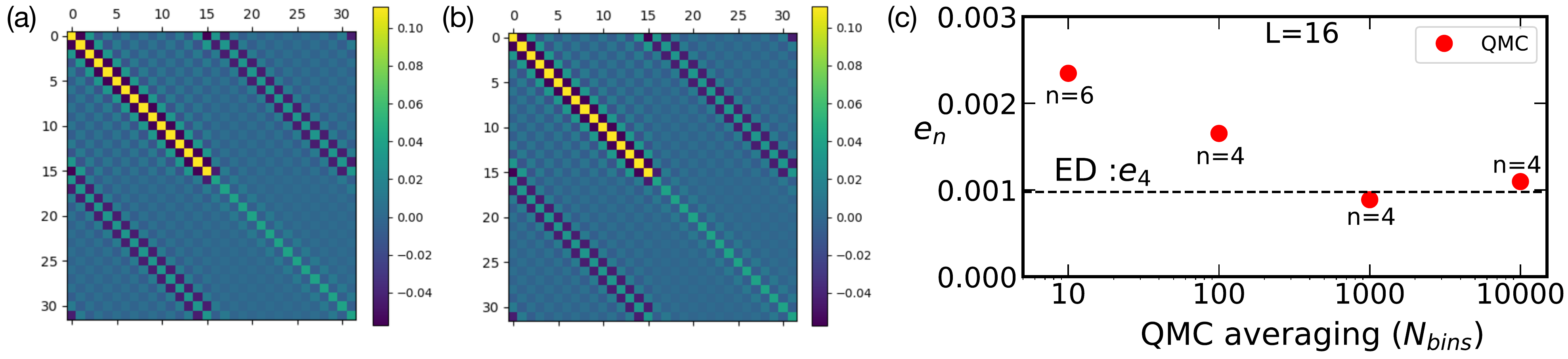}
}
\caption{(a)-(b)~We show the covariance matrix on a $L=16$ site chain computed with exact diagonalisation (ED) method (panel-(a)), and quantum Monte-Carlo (QMC) method (panel-(b)). (c)~We show that the eigenvalue of interest computed with QMC converging to the ED value for sufficiently large QMC averaging and thus giving us the correct eigenvector and hence the scar model.    
} 
\label{fig_SM0}
\end{figure*}

While using the EHC method, we are mainly focused on the lowest eigenvalues of the covariance matrix to build our target Hamiltonians. For small system sizes ($L \leq 24$), we can compare the covariance matrix computed with ED and QMC methods. In figure \ref{fig_SM0}(a-b), we show the covariance matrix for $L=16$. It can be seen that the matrix elements of the covariance matrix computed with ED and QMC agree perfectly. Furthermore the covariance matrix exhibits translational invariance. While the covariance matrices evaluated with ED and QMC methods agree pretty well, we are interested in the eigenvalues and eigenvectors of this for our EHC approach. With ED, we find that the target Hamiltonian defined by the fourth eigenvector (non-degenerate eigenvector) of the covariance matrix exhibits quantum many-body scars. We systematically increase the QMC measurement parameters, $N_{bins}$ and $N_{meas}$ and observe that for sufficiently large QMC averaging ($N_{bins} \approx 10^3$, $N_{meas} \approx 10^4$) of the EHC correlators, the eigenvalues of the covariance matrix computed via QMC converges to the ED values. In figure \ref{fig_SM0}(c), we show the behavior of fourth eigenvalue of the covariance matrix, $e_4$ with increasing $N_{bins}$ for a fixed $N_{meas} = 10^4$. In our QMC simulations, we typically use $N_{bins} \approx 1-10 \times 10^3$ and $N_{meas} = 10^4$ for measurement of our observables.

With increasing system sizes, we have to rely on the QMC method only. We find that not all target Hamiltonians defined by the fourth eigenvector of the covariance matrix exhibit quantum many-body scars. In stead, the target Hamiltonian we select corresponds to the smallest nontrivial and nondegenerate eigenvalue of the covariance matrix. We have systematically observed that this choice yields a target Hamiltonian with scarring properties.

\subsection{MBL model}

The Hamiltonian of interest is the $S=1/2$ antiferromagnetic Heisenberg model with a random disorder in 2D,
\begin{align}
H=J\sum_{\langle i,j\rangle} {\bf S}_i\cdot {\bf S}_j + \sum_{i} h_i S_i^z,
\label{SM_eq1}
\end{align}

We choose $N+1$ EHC basis operators such that ${\mathcal O}_0=\sum_{\langle i,j\rangle} {\bf S}_i\cdot {\bf S}_j $ and ${\mathcal O}_i=S_i^z,\;\; i=1,\ldots,N$. Thus the dimension of the covariance matrix is $(N+1) \times (N+1)$.

In the above Hamiltonian, the Ising term ($S_i^z S_j^z$) and the magnetic field term ($h_i S_i^z$) are the diagonal terms ${\mathcal O}^d$, while the exchange term ($S_i^x S_j^x+S_i^y S_j^y$) is the off-diagonal term ${\mathcal O}^{od}$~\cite{SSE_rev_Sandvik}.
To compute the elements of the covariance matrix, we carry out the measurement of following types of terms.

{\bf ${\mathcal O}^d {\mathcal O}^d$ terms :}
As both of the terms belong to diagonal type operation, we can do direct measurement for every spin state along the non-empty operator string. 
\begin{align}
    \langle {\mathcal O}^{d1} {\mathcal O}^{d2} \rangle = \big \langle \frac{1}{N_H} \sum_{p=1}^{N_H} {\mathcal O}^{d1}_p {\mathcal O}^{d2}_p \big \rangle,
\end{align}
\noindent where $N_H$ is the total number of the non-empty operator string in each measuring step, $p$ is the slice index, and $\big \langle ... \big \rangle$ is the average of Monte Carlo steps (see chapter-5 of Ref~\cite{SSE_rev_Sandvik} for further details).
As the spin state only changes during the off-diagonal operation, we can boost the efficiency by bookkeeping spins on most of the sites.

{\bf ${\mathcal O}^{od} {\mathcal O}^{od}$ terms :}
Only the exchange-exchange term in ${\cal C}$ belongs to this category. We cannot directly measure the off-diagonal term from the spin state~\cite{Sandvik1997}.  Instead, we use the number of appearance of the consecutive operators along the operator string to estimate its value. 
\begin{align}
	\langle {\mathcal O}^{od1} {\mathcal O}^{od2} \rangle = \frac{1}{\beta^2} \langle  (N_H-1) N_{c}({\mathcal O}^{od1}, {\mathcal O}^{od2}) \rangle 
\end{align}
\noindent where $N_{c}$ is the number of consecutive appearances of ${\mathcal O}^{od1}$ and ${\mathcal O}^{od2}$ along the operator string in each Monte Carlo step.

{\bf ${\mathcal O}^{d} {\mathcal O}^{od}$ terms :}
To calculate the combination of both diagonal and off-diagonal terms, we can combine both mentioned technique. At the occasion that ${\mathcal O}^{od}$ appears, we measure the ${\mathcal O}^{d}$ using direct measurement on the spin state.
\begin{align}
	\langle {\mathcal O}^{d1} {\mathcal O}^{od2} \rangle = \frac{1}{\beta} \langle \sum_{{\mathcal O}_p={\mathcal O}^{od2}} {\mathcal O}^{d1}_p \rangle 
\end{align} 
\noindent where ${\mathcal O}_p={\mathcal O}^{od2}$ is the slice that the operator is off-diagonal.

\section{Computation of the mean level spacing with QMC}
\label{appendix_C}

Working with a given system size (say $L \times L$), for a given disorder realization in $H$, 
we obtain the ground state $| \psi\rangle$ and built the new Hamiltonian $\tilde{H}$ via the eigenstate-to-Hamiltonian approach.
Then we run our QMC simulation on $\tilde{H}$ to compute the ground state energy of $\tilde{H}$ denoted by $E_{min}$. We also run QMC simulation on ${\cal -\tilde{H}}$, to compute the corresponding ground state energy, denoted as $-E_{max}$. From this, the mean level spacing is computed as $\Delta_N = \frac{E_{max} - E_{min}}{N}$, where $N$ is the number of states in the Hilbert space of $\tilde{H}$. Further averaging over disorder realisations is needed to obtain the average mean level-spacing.

\section{Computation of the variance of energy within the EHC approach}
\label{appendix_D}

We begin with the Hamiltonian,
\begin{equation}
H = J\sum_{\langle i,j\rangle} {\bf S}_i\cdot {\bf S}_j + \sum_{i} h_i S_i^z
\label{eq01}
\end{equation}
and obtain the ground state,
$$H | \Psi \rangle = E_0 | \Psi \rangle $$

We define the matrix elements of the covariance matrix,
$$
{\mathcal C}_{ij}=\langle {\mathcal O}_i {\mathcal O}_j \rangle -\langle {\mathcal O}_i \rangle \langle {\mathcal O}_j\rangle,
$$
where ${\mathcal O}_0=\sum_{\langle i,j\rangle} S_i^z S_j^z $ and 
${\mathcal O}_i=S_i^z,\;\; i=1,\ldots,N$. \\


We diagonalize the covariance matrix, and compute it's eigenvalues and eigenvectors. We find $e_1 = e_2 =0$, and $e_3 \neq 0$. The eigenvector corresponding to eigenvalue $e_3$, $\Psi_3 = \{\tilde{J}, \tilde{h}_1,\ldots,\tilde{h}_N \}$ 
defines a target Hamiltonian, $\tilde{H}$ such that 

\begin{equation}
\tilde{H} = \tilde{J} \sum_{\langle i,j\rangle} {\bf S}_i\cdot {\bf S}_j + \sum_{i} \tilde{h}_i S_i^z
\label{eq02}
\end{equation}

We rescale 
$\tilde{J} \rightarrow J$, $\tilde{h}_i \rightarrow (J/\tilde{J})\tilde{h}_i$, necessary for our QMC simulation, such that

\begin{equation}
\tilde{H} = J \sum_{\langle i,j\rangle} {\bf S}_i\cdot {\bf S}_j + \sum_{i} \tilde{h}_i S_i^z
\label{eq03}
\end{equation}

To compute the energy variance, we proceed as follows,


$$\tilde{H} | \Psi \rangle = H | \Psi \rangle +
(\tilde{H} - H) | \Psi \rangle  $$

$$\sigma^2(\tilde{H}) = \langle \Psi| \tilde{H}^2 | \Psi \rangle 
-(\langle \Psi| \tilde{H} | \Psi \rangle)^2 $$

\begin{align*}
\sigma^2(\tilde{H} - H)  &=  \langle \Psi| (\tilde{H} - H)^2 | \Psi \rangle 
-(\langle \Psi| (\tilde{H} - H) | \Psi \rangle)^2 \\
&= \langle \Psi| (\tilde{H}^2 - \tilde{H}H - H \tilde{H} + H^2) | \Psi \rangle \\
& ~~~~ -(\langle \Psi| \tilde{H} |\Psi \rangle - \langle \Psi| H | \Psi \rangle)^2 \\
&= \langle \Psi| \tilde{H}^2 |\Psi \rangle  - 2  \langle \Psi| \tilde{H}H |\Psi \rangle  +  \langle \Psi| H^2 | \Psi \rangle \\
& ~~~~ -(\langle \Psi| \tilde{H}| \Psi \rangle)^2  - 2\langle \Psi| \tilde{H} |\Psi \rangle \langle \Psi| H | \Psi \rangle  \\
& ~~~~ + (\langle \Psi| H | \Psi \rangle)^2 \\
& = \langle \Psi| \tilde{H}^2 |\Psi \rangle  -(\langle \Psi| \tilde{H}| \Psi \rangle)^2  - 2  \langle \Psi| \tilde{H}H |\Psi \rangle  \\
& ~~~~ + 2\langle \Psi| \tilde{H} |\Psi \rangle \langle \Psi| H | \Psi \rangle  +  \langle \Psi| H^2 | \Psi \rangle \\
& ~~~~ - (\langle \Psi| H | \Psi \rangle)^2 \\
& = \langle \Psi| \tilde{H}^2 |\Psi \rangle  -(\langle \Psi| \tilde{H}| \Psi \rangle)^2  \\
& ~~~~ +  \langle \Psi| H^2 | \Psi \rangle  - (\langle \Psi| H | \Psi \rangle)^2
\end{align*}

$$  \langle \Psi| H^2 | \Psi \rangle  - (\langle \Psi| H | \Psi \rangle)^2 = 0$$

\begin{eqnarray}
e_3 = \sigma^2(\tilde{H} - H) = \sigma^2(\tilde{H})
\label{eq04}
\end{eqnarray}

\eqref{eq04} is verified by using \eqref{eq01} and \eqref{eq02}. Furthermore, with use of \eqref{eq03} (instead of \eqref{eq02}), ~\eqref{eq04} changes to

\begin{eqnarray}
e_3/\tilde{J}^2 = \sigma^2(\tilde{H} - H) = \sigma^2(\tilde{H}) 
\end{eqnarray}

\section{MBL : Ground state phase transition and physical observables}
\label{appendix_F}

The ground state of $H$,~\eqref{SM_eq1}, has two distinct phases as disorder strength $h$ varies, with a quantum phase transition at a critical $h_c$. These phases may be characterised by measuring the spin stiffness, $\rho_s=\frac{1}{N}\frac{\partial^2E}{\partial \phi^2}$, defined as the response of the total energy, $E$, to a twist by angle $\phi$. 
The delocalized superfluid (SF) phase (for $h < h_c$) has finite spin stiffness, whereas the localized Bose glass (BG) phase (for $h > h_c$) has vanishing spin stiffness, and $h_c$, can be determined from the scaling of $\rho_s$.

In SSE, the stiffness is measured by the fluctuation in winding number($W$)
of the world lines as
$\rho_s = \langle W^2 \rangle/2\beta$, where $\beta$ is the inverse temperature~\cite{SSE_rev_Sandvik}.
Close to the critical point, the stiffness obeys the scaling relation
\begin{equation}
\rho_s(L,h) = L^{-z}f[(h-h_c)L^{1/\nu}],
\end{equation}
where 
the correlation length exponent is $\nu =1$~\cite{Prokofev2004}, and
the dynamical critical exponent is found to be $z = 2$.
Plotting the scaled stiffness $L^z\rho_s$ against $h$ for different system 
sizes provides an accurate estimate of the critical disorder strength, $h_c$~\cite{SF_BG_2D_HCB}.
The results are shown in Fig.~\ref{fig_SM3}(a), which suggest $h_c \approx 2.35$.
The interacting ground state changes from 
a delocalized superfluid state to a localized Bose glass state for $h > h_c$. 

\begin{figure*}[t]
\setcounter{figure}{1}
\renewcommand{\thefigure}{A\arabic{figure}}
\centerline{
\includegraphics[width=0.9\linewidth]{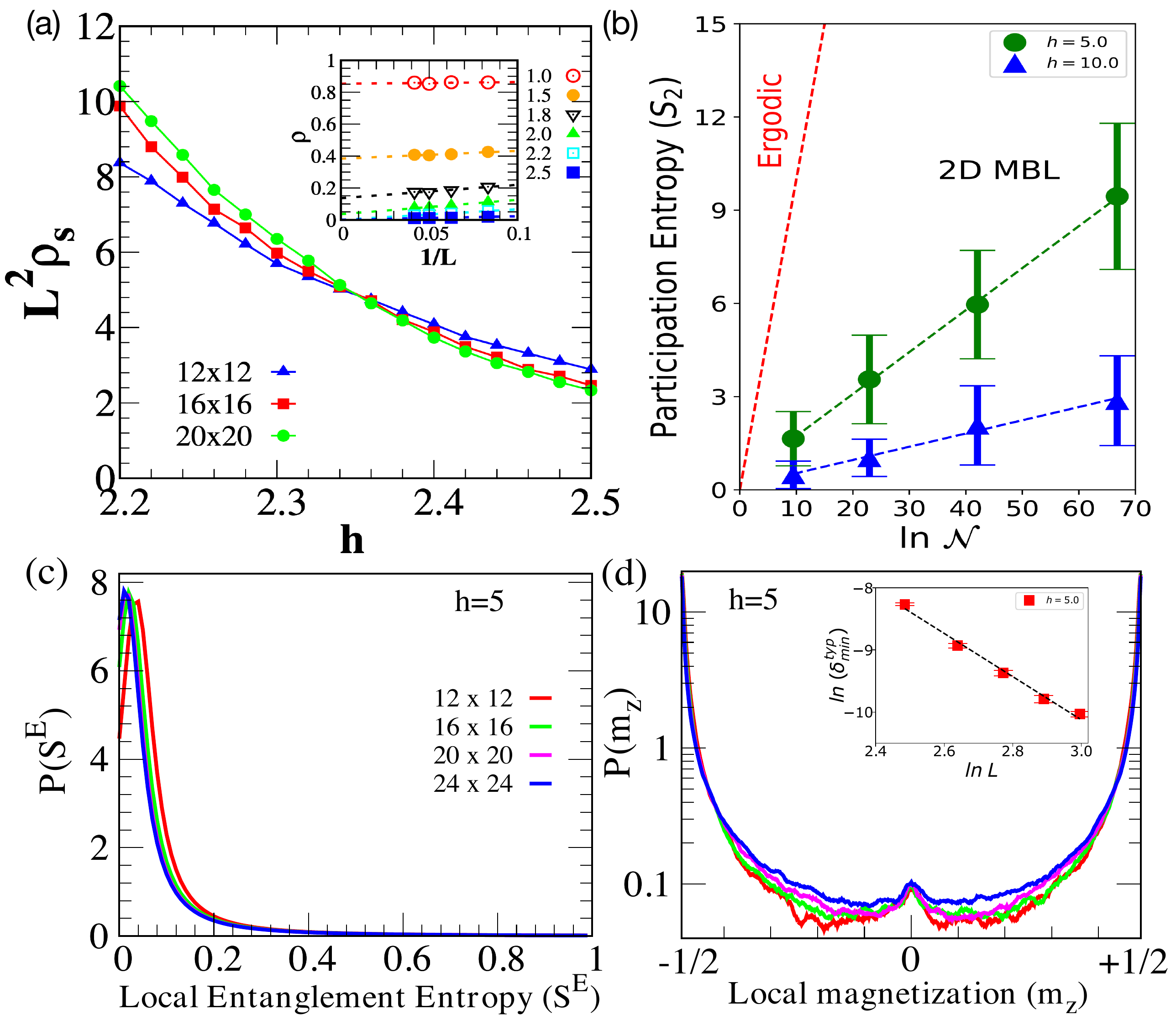}
}
\caption{
(a)~Behaviour of the scaled stiffness, $L^2\rho_s$, with varying $h$ near the transition region. The curves for different system sizes cross at $h=h_c$, providing an accurate estimate of the critical disorder strength, $h_c \approx 2.35$. (Inset)~Finite size scaling of the spin stiffness, $\rho_s$, with varying system sizes for different disorder strengths. In the thermodynamic limit, $\rho_s \to 0$ as $h \geq h_c$ increases, establishing the BG phase as the ground state.
(b)~Scaling of second order R\'{e}nyi entropy, $S_2$ with the Hilbert space size ${\cal N}$ in the presence of disorder, demonstrating the non-ergodic behavior of the Bose glass ground state.
(c)~Distribution of local entanglement entropy $P(S^E)$ in the ground state for various system sizes for $h=5$. $P(S^E)$ shows a sharp peak at $S^E \sim 0$ indicating that each site is almost disentangled from the other sites, a characteristic signature of MBL~\cite{Wahl2019}. As expected, the $S^E$ peak moves towards $S^E = 0$ with increasing system sizes.
(d)~(Main panel)~Distribution of local magnetization $P(m_z)$ in the ground state for different system sizes for $h=5$. $P(m_z)$ is strongly peaked at the values $m_z = \pm 1/2$, indicative of the local moments being fully aligned with the local random magnetic field. 
(Inset)~Power-law decay of maximum polarization $\delta_{min}$ (see text) with system size, another characteristic signature of MBL~\cite{Dupont2019b,KT_MBL_Gabriel2020}.
} 
\label{fig_SM3}
\end{figure*}

{\bf Participation Entropy :}
The $q$-th order R\'{e}nyi participation entropy of a state $\ket{\psi}$ is given by
\begin{equation}
S_q = \frac{1}{1-q}\ln \sum_i p^q_i,
\end{equation}
where $p_i=|\braket{\psi|\phi_i}|^2$ and the $\ket{\phi_i}$ are some set of orthonormal basis states.
In particular, we focus on $q=2$ and $q\to\infty$. These two quantities provide the measure of how many states of a configuration space contribute to a wave function.

We use the approaches developed in \cite{Renyi_entropy_Roscilde_2012,Luitz2014} to calculate the participation entropy. These approaches use the counting of occurrence for each spin configuration to calculate the participation entropy.
$S_q$ is found using the probability of having identical configurations in different replica in each Monte Carlo step, while $S_\infty$ is calculated using the probability of maximally occurring spin configuration. For strong disorders, the maximally occurred spin configuration is usually almost aligned with the local magnetic field.

In Fig.~\ref{fig_SM3}(b), we show the scaling of disorder-averaged $S_2$ with the Hilbert space size, ${\cal N}$ in the localised regime. The slope of the line $S_2 = D_2 \ln{\cal N} + c$ represents the multifractal dimension $D_2$, and we find $D_2 \ll 1$. This indicates that only a vanishingly small fraction of basis states (among the exponentially large space of states in the configuration space) contribute to the Bose glass ground state in our simulations; highlighting it's strong non-ergodic behavior. The behavior of $S_\infty$ is shown in the main text.

{\bf Local entanglement entropy :} 
We measure the local entanglement entropy 
$S^E = - \ln  \text{Tr} \rho_{loc}^2$
for a bipartition of the system where the subsystem of interest is chosen to be one site only, using the SSE extended ensemble scheme~\cite{Renyi_entropy_Roscilde_2012,Luitz2014}.
In Fig.~\ref{fig_SM3}(c), the distribution of $S^E$, $P(S^E)$ shows a sharp peak close to $S^E = 0$.
This is a prominent feature of MBL (see Ref.~\cite{Wahl2019}), where any given site is almost disentangled from other sites of the lattice and its reduced density matrix, $\rho_{\text{loc}}$ can be approximated as that of a pure state.
Furthermore, our results show a convergence of the data with different system sizes.

{\bf Local magnetization :}
We study the distribution of local magnetization $P(m_z)$ in Fig.~\ref{fig_SM3}(d), and find a bipolar distribution with peak values at $m_z = \pm 1/2$, a signature of polarization along the on-site disordered magnetic field.
Following Refs.~\cite{Dupont2019a, KT_MBL_Gabriel2020}, we further look into the maximum polarisation, defined as $\delta_{\rm min} = 1/2 - {\rm max}(|m_z^i|)$. We observe that the typical average of $\delta_{\rm min}$, $\delta_{\rm min}^{\rm typ} \propto L^{-\gamma}$, with $\gamma \sim 3.5$ for $h=5$ (see inset). This behavior is analogous to the freezing of local moments in the MBL phase~\cite{Dupont2019a, KT_MBL_Gabriel2020}.

\bibliography{references}

\end{document}